\documentclass[]{tPHM2e}

\usepackage[latin1]{inputenc} 
\usepackage{graphicx,wrapfig}
\usepackage{amsmath,amssymb}
\usepackage{xspace}

\newcommand{\etal}{\textit{et al}.\@\xspace}
\newcommand{\ie}{\textit{i.e.\@\xspace}}
\newcommand{\cf}{\textit{cf\@\xspace}}
\newcommand{\md}{\mathrm{d}}
\DeclareMathOperator{\trace}{Tr}

\begin{document} 
\doi{10.1080/14786435.20xx.xxxxxx}
\issn{1478-6443}
\issnp{1478-6435}
\jvol{00} \jnum{00} \jyear{2010} %\jmonth{21 December}

\markboth{B. Bak{\'o} et al.}{Philosophical Magazine}

%\articletype{?}

%<<< Title, author, ...
\title{Dislocation dynamics simulations with climb:\\ 
kinetics of dislocation loop coarsening controlled by bulk diffusion}

\author{Botond Bak\'o$^{\rm{a}}$,
Emmanuel Clouet$^{\rm{a}}$$^{\ast}$
\thanks{$^\ast$Corresponding author. Email: emmanuel.clouet@cea.fr\vspace{6pt}},
Laurent M. Dupuy$^{\rm{b}}$
and 
Marc Bl\'etry$^{\rm{c}}$
\\\vspace{6pt}  
$^{\rm a}$ 
\emph{CEA, DEN, Service de Recherches de M\'etallurgie Physique, \\
91191 Gif-sur-Yvette, France} \\
$^{\rm b}$ 
\emph{CEA, DEN, Service de Recherches de M\'etallurgie Appliqu\'ee, \\
91191 Gif-sur-Yvette, France} \\
$^{\rm c}$ 
\emph{Institut de Chimie et des Mat\'eriaux Paris-Est, CNRS UMR 7182,\\
2-8, rue Henri Dunant, 94320 Thiais, France}
}

\maketitle
%>>>
\begin{abstract}
	Dislocation climb mobilities, assuming vacancy bulk diffusion,
	are derived and implemented in dislocation dynamics simulations
	to study the coarsening of vacancy prismatic loops in fcc metals.
	When loops cannot glide, the comparison of the simulations
	with a coarsening model based on the line tension approximation
	shows a good agreement.
	Dislocation dynamics simulations with 
	both glide and climb are then performed.
	Allowing for glide of the loops along their prismatic cylinders
	leads to faster coarsening kinetics, as direct coalescence
	of the loops is now possible.
	
	\bigskip


\end{abstract}

\section{Introduction}

The strength of crystalline materials is mainly determined by the motion 
of dislocations, the carriers of plastic flow. At high temperatures the 
mechanical properties of metals and alloys can change fundamentally because
of dislocation climb. Climb occurs when the motion of the dislocations has a 
component perpendicular to their glide plane. 
This requires the emission or absorption of point defects
and their long-range diffusion. 
Due to the ability to annihilate edge 
dislocation dipoles, climb plays a fundamental role, for instance, in high temperature 
creep \cite{creep,POI85} and recovery \cite{humphreys}.
Climb mobility has also an important effect on dislocation morphology 
\cite{rudolph1,rudolph2}. 
Specifically, climb and cross-slip control the length scale of the cells
of the dislocation network \cite{bako1,bako3,bako2},
and the suppression of climb leads to the freezing of the network
into a diffuse-looking random distribution \cite{bako2,bako4}.
It is thus highly desirable to incorporate these mechanisms in dislocation
dynamics (DD) simulations which constitute the most suitable tool to study
the evolution of a whole dislocation population, and thus to model the 
plastic flow at a mesoscopic scale.

Dislocation climb was introduced in several two- 
\cite{roters1,bako1,bako3,bako2} and three-dimensional (3D) 
\cite{cai1,xiang,chen2010,ghoniem2000,mordehai2008,mordehai2009} DD simulations. 
These models generally treat dislocation climb as a glide motion,
\ie a conservative motion, with a smaller mobility.
This is not sufficient to capture all the involved physics.
When the point defect concentration in the bulk is different from its 
equilibrium value (e.g. under irradiation or after a quench of the crystal),
dislocations can climb without the existence of a mechanical force.
Point defect super- or under-saturation gives birth, in this case,
to an additional force, the osmotic force.
This is not described in simulations where climb is treated like glide,
and only DD models based on the diffusion theory of point defects
\cite{turunen1976,ghoniem2000,raabe,mordehai2008,mordehai2009}
lead then to a proper description of dislocation climb.

We introduce in the present article a diffusion based climb model 
in 3D DD simulations using a nodal representation of dislocations
\cite{bulatov2006,arsenlis2007}. The climb model is similar to the one previously 
used by Mordehai \etal \cite{mordehai2008,mordehai2009} in 3D DD simulations
where dislocations were discretized in edge and screw segments.
These DD simulations are used to study the coarsening kinetics 
of prismatic dislocation loops.

Prismatic dislocation loops, whose Burgers vector has a component normal 
to their habit plane, may be obtained in both thermal quenching 
\cite{hirsch58,silcox1960} and irradiation 
experiments \cite{masters1965,eyre1971,eyre1973,kawanishi1986}.
They result from the condensation of point-defects, 
vacancies or interstitials, into discs which collapse to form 
a dislocation loop.
These loops contribute in irradiated metals 
to the material embrittlement by radiation damage.
In semiconductors, loops formed during the annealing stage
which follows ion implantation may affect electronic properties of the device
\cite{jones1988,bonafos1998,persson2006}.
Knowing the long term time evolution of the loop population is therefore
essential.

It has been observed experimentally 
\cite{johnson1960,silcox1960,eyre1971,jones,liu1995,bonafos1998,persson2006}
that loops coarsen: large loops grow at the expense of 
the smaller ones. Their size thus increases on average,
whereas their density decreases.
Different mechanisms for loop coarsening have been proposed 
and modeled. 
Large loops may grow by the absorption of vacancies 
emitted by the shrinking loops, the exchange being carried out 
through vacancy bulk diffusion 
\cite{silcox1960,eyre1971,kirchner1973,powell1975,burton1986,enomoto1989,liu1995,bonafos1998}.
Pipe diffusion of point defects along the dislocation lines
may also lead to a transfer of matter around the loops.
This generates a translation of the loop in its habit plane,
a process known as self climb or conservative climb \cite{hull2001}.
Coarsening can then occur by direct coalescence of the loops
\cite{johnson1960,eyre1971,persson2006}.
Finally, if loops are unfaulted, they can glide along their prismatic
cylinder, thus also allowing for coalescence \cite{eyre1971}.
Depending of the material, the type of the loops (vacancy or interstitial,
faulted or unfaulted), and the annealing temperature, loop coarsening
occurs by one or several of the mechanisms described above.
In the present article, we do not consider pipe diffusion, 
which will be devoted to future work,
and we study loop coarsening controlled by bulk diffusion.
This coarsening regime has been reported in several experiments
\cite{powell1975,liu1995,bonafos1998}.

The paper is organized as follows. 
In the next section we describe our climb model and its introduction
in 3D DD simulation. 
Then simulations of loop coarsening are performed, where  prismatic
glide of the loop is forbidden.
This allows a comparison with the coarsening model first proposed by Kirchner 
\cite{kirchner1973}, and revisited by Burton and Speight \cite{burton1986}.
Finally, the contribution of prismatic glide to the kinetics is presented.
The paper ends with concluding remarks.

\section{Dislocation climb model}

The climb model used in our DD simulations is based on the
diffusion theory of point defects. 
As our DD simulations are used to study the annealing
kinetics of post-irradiated or quenched materials,
only vacancies are contributing to dislocation climb.
Interstitials play a role only under irradiation.
Jogs and pipe-diffusion are not considered in our model.
This means that dislocations are assumed to be perfect sinks for vacancies
which are thus at equilibrium all along the dislocation lines
and not only on localized points corresponding to the jogs.
These are the same assumptions as in the work of Mordehai 
\etal \cite{mordehai2008},
where it has been shown that such an idealized model leads 
nevertheless to a reasonable description of dislocation climb
and is justified for high temperatures like the one of the present work (600\,K in Al). 
In contrast with this earlier work, our DD simulations are based
on a nodal representation of the dislocation line
and not on a discretization in edge and screw segments.
We therefore avoid discretization problems\footnote{In their work,
Mordehai \etal needed to correct the climb velocity of the edge segments
to consider the actual orientation of the dislocation line (\S~2.3 in Ref. \cite{mordehai2008}).},
but we need to define a climb velocity for all the dislocation
characters, and not only for the edge ones.

\subsection{Climb mobility law}

The starting point to derive the dislocation climb rate 
is the diffusion equation for the vacancy concentration $c$ 
in the steady-state limit.
Neglecting the elastic interaction between vacancies and dislocations,
this reduces to the Laplace equation,
\begin{eqnarray}
\Delta c({\bf r}) = 0.
\label{eq:laplacecv}
\end{eqnarray} 
In order to obtain simple analytical expressions, 
we derive the solution of this equation for 
an isolated infinite straight dislocation. 
We therefore do not consider the interaction between 
the diffusion fields of the individual dislocation segments.

A cylindrical control volume with inner radius $r_{\rm c}$ of order of 
the core radius is defined around the dislocation segment whose climb rate we want 
to calculate. 
As we are not taking into account jogs nor pipe-diffusion, 
vacancies which diffuse into this control volume 
are absorbed immediately.
At the distance $r_{\rm c}$ from the line, they are at equilibrium with the 
dislocation. This leads to a vacancy concentration \cite{friedel1964,turunen1976}
\begin{eqnarray}
	c_{\rm eq} = c_0\exp{ \left (\frac{F_{\rm cl}\Omega}
	{kT b \sin{(\theta)} } \right ) } ,
	\label{cvac}
\end{eqnarray} 
where $c_0 = \exp{[-(U^{\rm f}_{\rm v}-P\Delta V_{\rm v})/kT]}$ is the equilibrium vacancy 
concentration in the defect-free crystal at the pressure $P$, 
$\Omega$ is the atomic volume, 
$\theta$ describes the dislocation character\footnote{$\theta$ is defined
from $\sin{(\theta)}=\lVert\mathbf{b}\times\mathbf{\zeta}\rVert/b$,
as a consequence, $\sin{(\theta)}$ is always positive.},
\ie the angle between its line direction unit vector
$\mathbf{\zeta}$ and its Burgers vector $\mathbf{b}$, 
$U^{\rm f}_{\rm v}$ is the vacancy formation energy
and $\Delta V_{\rm v}$ the associated relaxation volume\footnote{The 
pressure dependence \cite{weertman1965,lothe1967}
of the equilibrium vacancy concentration 
is not considered in the present simulations where a relaxation volume 
$\Delta V_{\rm v}=0$ is assumed.}, 
$k$ is the Boltzman's constant 
and $T$ is the temperature. 
The mechanical climb force $F_{\rm cl}$ is the projection
of the Peach-Koehler force in the direction perpendicular to the 
dislocation glide plane \cite{lothe1967},
\begin{eqnarray}
	F_{\rm cl} = \left[ (\sigma {\bf b})\times{\bf \zeta} \right] 
	\cdot \mathbf{n},
	\label{eq:Fcl}
\end{eqnarray} 
where $\sigma$ represents the stress tensor acting on the dislocation 
segment. It is thus the combination of the stress
created by all other segments present in the simulation 
and of the externally applied load.
The hydrostatic part of this tensor, $P=-1/3\trace{(\sigma)}$,
gives the pressure controlling 
the vacancy equilibrium concentration in Eq.~(\ref{cvac}).
The normal $\mathbf{n}$ to the dislocation glide plane appearing 
in Eq. (\ref{eq:Fcl}) is defined for non screw dislocation
by the convention
\begin{equation}
	\mathbf{n} = \frac{ \mathbf{b} \times \mathbf{\zeta} }
	{ \lVert \mathbf{b} \times \mathbf{\zeta} \rVert}
	= \frac{ \mathbf{b} \times \mathbf{\zeta} }{ b \sin{(\theta)} }.
	\label{eq:normal}
\end{equation}
With such a convention, a dislocation emits vacancies when it climbs 
in the direction of $\mathbf{n}$, and absorbs vacancies otherwise.

The solution of Eq. (\ref{eq:laplacecv}) at distance $r$ from the 
dislocation segment is  obtained by imposing $c(r = r_\infty) = 
c_\infty$ in the bulk, far from the dislocation, leading to 
\begin{eqnarray}
c(r) = c_\infty + (c_\infty - c_{\rm eq})\frac{\ln (r/r_\infty)}{\ln(r_\infty/r_{\rm c})}.
\end{eqnarray} 

An infinitesimal dislocation segment of length $\delta l$
emits vacancies with a rate given by 
\begin{eqnarray}
        \delta N_{\rm v} 
        & = & -2\pi r_{\rm c} \delta l 
	\frac{D_{\rm v}}{\Omega} \left. \frac{\partial c}{\partial r} \right|_{r=r_{\rm c}} \nonumber\\
	& = & - 2\pi \delta l \frac{D_{\rm v}}{\Omega} 
         \frac{c_\infty - c_{\rm eq}}{\ln (r_\infty/r_{\rm c})},
\end{eqnarray} 
where the vacancy bulk diffusion coefficient is given by
\begin{eqnarray}
D_{\rm v} = D_{\rm v}^0 \exp\left( -\frac{U_{\rm v}^{\rm d}}{kT}\right)
\end{eqnarray}
with $U^{\rm d}_{\rm v}$ being the vacancy migration energy and 
$D_{\rm v}^0$ a constant pre-factor characterizing vacancy diffusion.
If $\delta N_{\rm v} < 0$, the segment actually absorbs vacancy.
Its climb velocity $\mathbf{v}_{\rm cl} = v_{\rm cl}\mathbf{n}$ 
is given by
\begin{eqnarray}
	v_{\rm cl} 
	&=& \delta N_{\rm v} \frac{ \Omega }
		{b \delta l \sin{(\theta)} }\nonumber\\
	& = & \frac{2\pi D_{\rm v}c_0}{b \sin{(\theta)} \ln{(r_\infty/r_{\rm c})}}
		\left [\frac{c_{\rm eq}}{c_0} - \frac{c_\infty}{c_0}\right].
	\label{vclimb}
\end{eqnarray} 

Eq. (\ref{vclimb}) actually gives the climb rate of the infinitesimal 
dislocation segment of length $\delta l$. We need to deduce from it 
the velocity of the nodes. 
This is done using ``shape functions'' \cite{bulatov2006,arsenlis2007}, \ie functions
which are non-zero only when a spatial point lies on the segment connected to
a given node.
This allows to define nodal forces by integration along each segment 
of the forces acting on this segment. 
This integration is done using 5 Gauss points on each segment
with weights given by the shape function. 
The nodal velocities are obtained by solving the set of equations 
that link the nodal forces to the nodal velocities
\cite{bulatov2006,arsenlis2007},
using mobility laws such as the one given by Eq. (\ref{vclimb}).
This is more easily done if the velocity varies linearly with 
the applied force. We therefore linearized Eq. (\ref{vclimb})
by taking advantage that the climb force $F_{\rm cl}$ is small enough
so that the exponential appearing in Eq. (\ref{cvac}) can be
expanded to the first order. This leads to the linear relation
\begin{eqnarray}
v_{\rm cl} = M_{\rm cl}[F_{\rm cl} + F_{\rm os}],
\label{vclimb1}
\end{eqnarray} 
where we have defined the climb mobility
\begin{eqnarray}
	M_{\rm cl}(\theta) 
	= \frac{2\pi D_{\rm v}\Omega c_0}
	{kTb^2\sin^2{(\theta)}\ln(r_\infty/r_{\rm c})},
	\label{eq:clmob}
\end{eqnarray} 
and the osmotic force \cite{friedel1964,weertman1965,lothe1967} 
\begin{eqnarray}
	{\bf F}_{\rm os}(\theta) = \frac{kT}{\Omega}
	\left ( 1 - \frac{c_\infty}{c_0}\right )
	b \sin{(\theta)} \mathbf{n}.
	\label{eq:Fos}
\end{eqnarray} 
With this force, dislocations are climbing in presence of a vacancy 
supersaturation.

Considering that each point along the dislocation line may act 
as a sink or source of vacancies, and thus neglecting the effect
of jogs and pipe-diffusion, implies that the climb mobility and the 
osmotic force depend only on the dislocation character, $\theta$.
As seen from Eq. (\ref{eq:Fos}), the osmotic force tends to zero
for a dislocation of screw character: a point defect
supersaturation does not exert any force on a pure screw
dislocation.
But the climb mobility given by Eq. (\ref{eq:clmob}) is diverging
for a screw dislocation. If the Peach-Koehler force has a component
perpendicular to the dislocation glide plane\footnote{As we are modeling
climb in fcc crystals where dislocations are dissociated,
the glide plane can be defined even for a screw dislocation 
although $\protect\mathbf{b}\times\protect\mathbf{\zeta}=\protect\mathbf{0}$ in this case.}, 
our model leads to an infinite climb velocity for a screw dislocation.
This artifact is usual in models where a climb mobility is defined 
for all dislocation characters \cite{turunen1976} without incorporating
jogs in the modeling. A proper account of the interaction between
a vacancy and a jog in the case of a screw orientation, may remove
the divergence of the climb mobility.
In the present work, the divergence is handled by considering
that screw dislocations do not climb. We therefore enforce zero
climb velocity for all segments which are nearly parallel with their
Burgers vector. When the dislocation character $\theta$
is less than $10^{-6}$\,radians, the segment is handled in our code
as pure screw. We check that varying this threshold does not change 
the results of our simulations.

The incorporation in the diffusion equation (\ref{eq:laplacecv}) of the elastic interaction 
between vacancies and dislocations will change the climb mobility (\ref{eq:clmob})
by multiplying it with a prefactor depending on the interaction energy and the temperature 
\cite{Wolfer2007}. It will therefore only lead to a correction on the time scale. 
Such an elastic interaction is important to consider when two different point defects, 
like vacancies and interstitials, are diffusing as it can lead to some bias on their relative 
absorption by dislocations \cite{Wolfer2007}. In our simulations, only vacancies are present 
and we therefore neglect the effect of this elastic interaction.

\subsection{Simulation setup}

For simulations the materials parameters of fcc aluminum are used: 
lattice constant $a=0.404$\,nm, shear modulus $\mu = 26.5$\,GPa, 
Poisson coefficient $\nu=0.345$,
vacancy migration energy $U_{\rm v}^{\rm d} = 0.61$\,eV, vacancy 
formation energy $U^{\rm f}_{\rm v}=0.67$\,eV, atomic volume $\Omega = 
16.3 \times 10^{-30}{\rm \,m}^3$, diffusion coefficient pre-exponential 
$D_0 = 1.18\times 10^{-5}{\rm \,m}^2{\rm s}^{-1}$.

At each simulation step, the stresses, forces, and velocities are 
calculated. No external loading is applied.
Then the dislocations are moved using an integration time 
interval inversely proportional to the maximum velocity of all nodes. 
The nodes are allowed to fly over a maximum distance of one Burgers 
vector in case of glide motion and $b/10$ 
in case of climb, this defining the duration of the time step for 
integration.

No image stress corresponding to free surfaces or periodic boundary 
conditions are considered. The stress field calculation therefore 
assumes that dislocations are in an infinite homogeneous elastic medium. 
The stress calculation is performed using the non singular expressions
of Cai \etal \cite{cai2006} assuming isotropic elasticity 
with a parameter $a=3$\,{\AA} for the core spreading.

For the numerical simulations of prismatic loop coarsening,
the loops are of vacancy type and we assume that they are 
far enough from surfaces and grain boundaries, so that
the loops are the only sources and sinks for vacancies.
The total number of vacancies in the system, \ie the sum 
of vacancies condensed in the loops and the free vacancies 
diffusing in the bulk, is therefore a conserved quantity.
Climbing dislocations emit or absorb a number of vacancies 
proportional to the area they sweep. The vacancies in the bulk are 
considered to reach steady state instantaneously. The time evolution of 
the bulk vacancy concentration is then governed by the equation
\begin{eqnarray}
\frac{{\rm d}c_\infty}{{\rm d}t} = \frac{b}{V}\frac{{\rm d} S}{{\rm d}t},
\end{eqnarray} 
where $V$ is the volume of the sample 
and $S$ is the area swept by the loops during climb.
If a dislocation segment $\delta l$ climbs a distance $v_{\rm cl}\ \Delta t$,
the corresponding swept area is given by 
$\delta S = v_{\rm cl}\ \Delta t\ \delta l \sin{(\theta)}$.
The sign of the climbing velocity fixes the sign of the swept area
and therefore determines if the dislocation segment is absorbing ($\delta S<0$)
or emitting vacancies ($\delta S>0$). 

\section{Climb-only controlled coarsening}

\subsection{DD simulations of loop coarsening}

We first use the climb model to simulate the coarsening kinetics of prismatic
loops which are not allowed to glide, thus  putting the glide mobility to zero
in the DD simulations.
This corresponds to the behavior of faulted loops which cannot glide 
on their prismatic cylinder because of the existing stacking fault.
The additional force exerted by the stacking fault on the dislocation
segments is not included in the simulation.

The initial configuration is a population of circular loops
which are placed at random in a box of size $L\times 
L\times L$, $L = 2$\,$\mu$m, and with the condition that they do not 
overlap and the disks corresponding to them do not intersect.
Their habit planes are also chosen at random from the plane family 
$\{110\}$.
The Burgers vector of the loops is of $1/2\ \langle110\rangle$ type
and is perpendicular to the loop habit plane.
The loops are then pure prismatic. 
As glide is not allowed, they can only grow or shrink in their habit plane
by climb. As a consequence, they remain pure prismatic during the 
whole simulations. This means that all the dislocation segments 
are pure edge.
These loops are of vacancy type and thus the glide plane normal 
$\mathbf{n}$, as defined by Eq. (\ref{eq:normal}), is pointing
at every point to the center of the loop.

\begin{figure}[!tbhp]
\begin{center}
\includegraphics[width=0.75\textwidth]{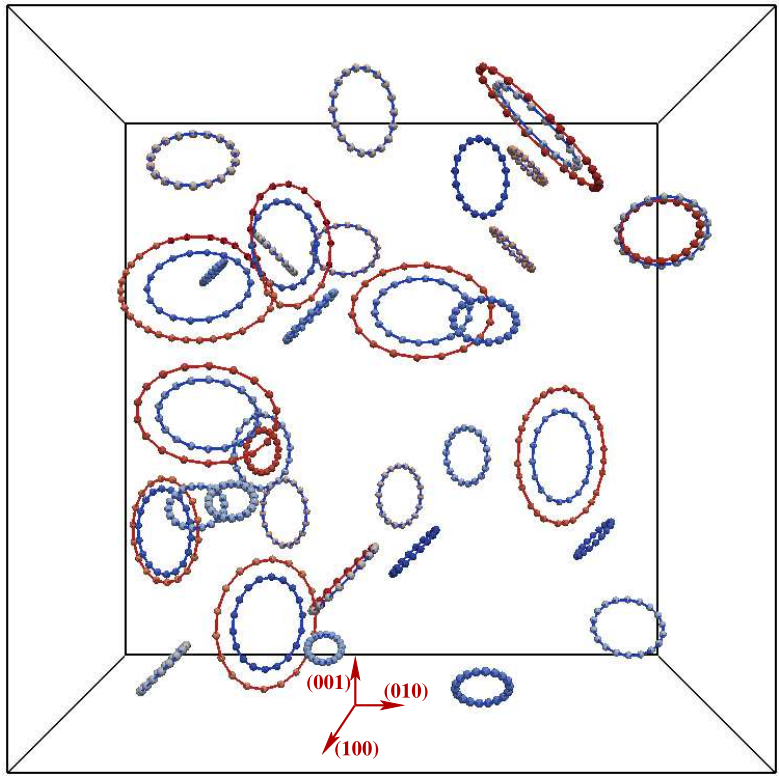}
\end{center}
\caption{[Color online] Typical configuration of loops, when glide is 
not possible. Darker [blue] lines represent the loops at $t = 0$
and lighter [red] lines at $t=1$\,s.
(See video as supplementary online material.)}
\label{circles}
\end{figure}

We set the temperature to a constant value, $T=600$\,K, 
during the simulation, and we start with no vacancy 
supersaturation: $c_{\infty}(t=0)=c_0$.
For a short time at the beginning, all loops 
start shrinking because of their line tension and emit vacancies.
As a result, the vacancy concentration $c_{\infty}$ in the bulk
increases.
This transient regime occurs because $c_0$ is the equilibrium
vacancy concentration in a defect-free crystal. 
Because of the line tension of the loops and of the associated
Gibbs-Thomson effect, the vacancy concentration in equilibrium
with the vacancy prismatic loops needs to be higher than $c_0$.
This vacancy supersaturation creates an osmotic force.
The coarsening kinetics will then result from the competition 
between this osmotic force and the mechanical climb force. 
For small loops, the line tension leads to a mechanical climb
force higher than the osmotic force, thus making the loops shrink. 
Larger loops have a smaller line tension. Therefore the osmotic 
force is higher than the mechanical climb force for them.
The larger loops grow then by absorbing the vacancies emitted by the smaller ones
which are shrinking. The frontier between stable and unstable
loops is controlled by the instantaneous vacancy concentration 
$c_{\infty}(t)$ through the Gibbs-Thomson effect. As $c_{\infty}(t)$
is tending to $c_0$ with the time evolution, this frontier is 
going to the larger sizes of the loops.
As a consequence, as time evolves, the loops grow on average in radius
whereas their density decreases. 
A typical configuration at $t = 0$ and time $t 
= 1$\,s is shown in Figure \ref{circles}. 

\subsection{Coarsening kinetics}

\begin{figure}
\begin{center}
\includegraphics[angle=270,width=0.75\textwidth]{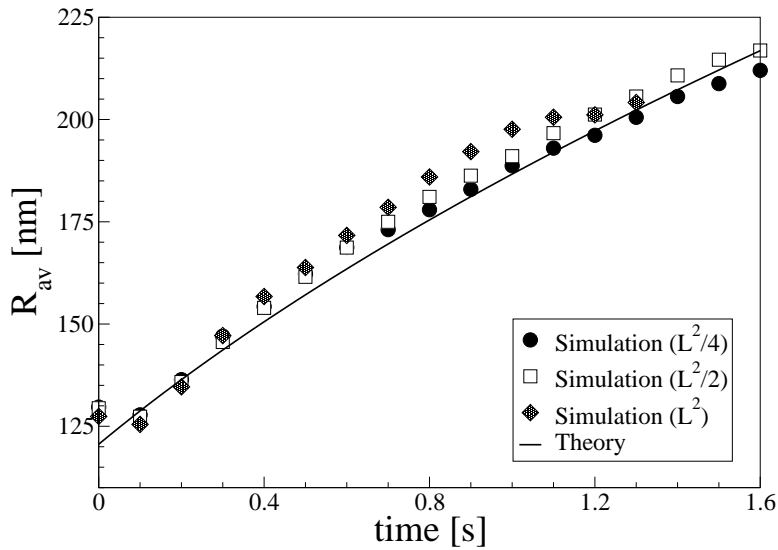}
\end{center}
\caption{Time evolution of the average radius of the loops for three different
initial total areas $L^2/4$, $L^2/2$, and $L^2$. The line corresponds to Eq. 
(\ref{ravt}).}
\label{r}
\end{figure}

\begin{figure}
\begin{center}
\includegraphics[angle=270,width=0.75\textwidth]{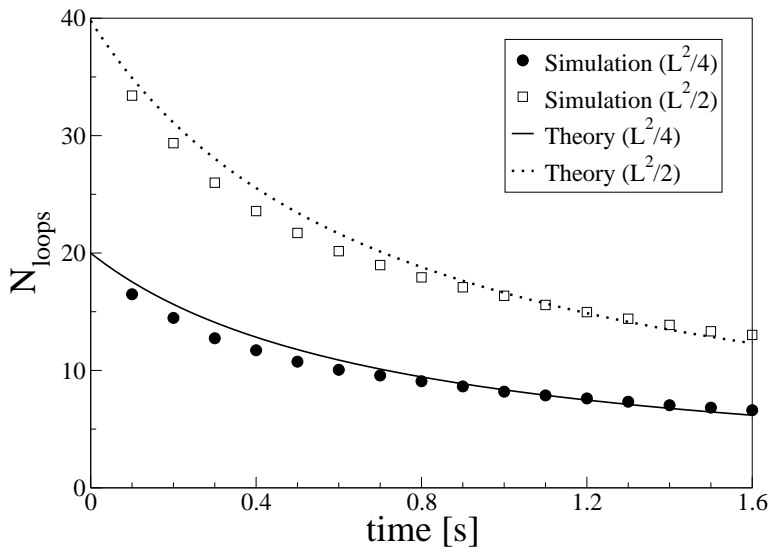}
\end{center}
\caption{Time evolution of the average number of loops for two 
different total initial areas $L^2/4$ and $L^2/2$. 
The lines correspond to Eq. (\ref{nt}).}
\label{fig5}
\end{figure}

\begin{figure}
\begin{center}
\includegraphics[width=0.75\textwidth]{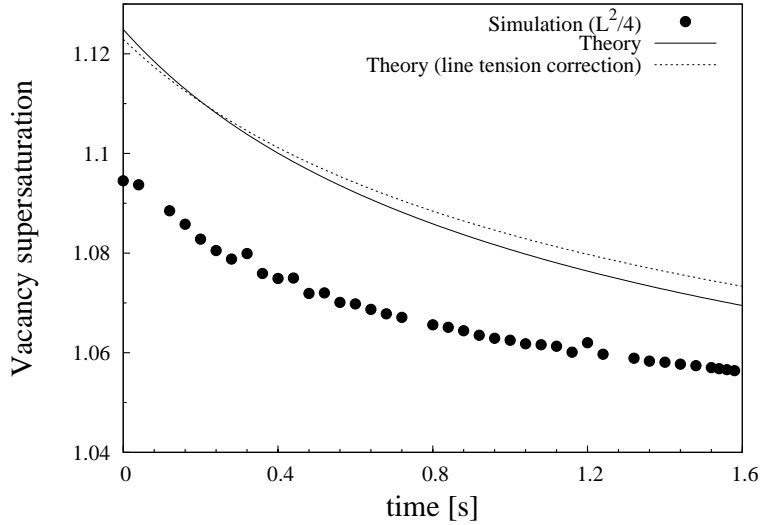}
\end{center}
\caption{Time evolution of the vacancy supersaturation, $c_\infty(t)/c_0$,  
for the initial total loops area $L^2/4$.
The lines are predictions of the KBS model for different
line tension approximations.
The solid line corresponds to Eq. (\ref{cc0}) where the line tension is $\mu b/R$.
The dashed line corresponds to Eq. (\ref{eq:cc0_KBS_improved}) where the line tension
is given by Eq. (\ref{eq:line_tension}) with $r_{\rm c}=1.5$\,\AA.}
\label{f}
\end{figure}

\begin{figure}
\begin{center}
\includegraphics[angle=270,width=0.75\textwidth]{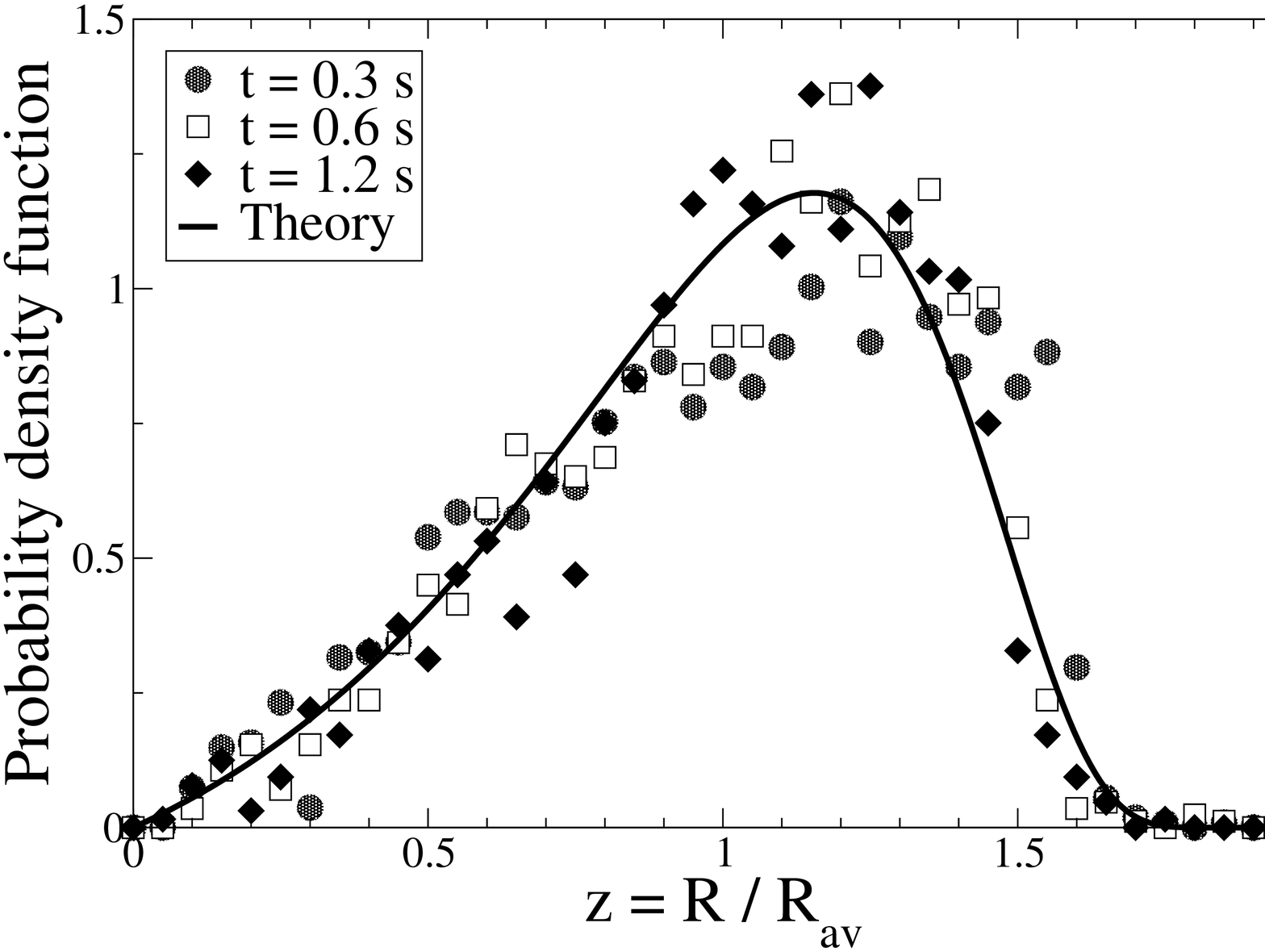}
\end{center}
\caption{Probability distribution function of the normalized loop sizes 
for the initial total loops area $L^2/4$.}
\label{figdistrib}
\end{figure}

To extract quantitative information from our DD simulations, 
we consider 500 statistically equivalent systems
with different realization of randomness.
The prismatic loops at $t=0$ have random radii generated with uniform 
distribution, with the constraint that their total area at $t=0$ is 
the same in all simulations. 
Simulations have been performed for three different initial 
total loop areas $S(0) = L^2/4$, $S(0) = L^2/2$, $S(0) = L^2$, 
where $L$ is the size of the simulation box.
The time evolution of the average radius $R_{\rm av}$ of the loops, 
of the average number of loops $N_{\rm loops}(t)$ in the simulation box,
and of the vacancy supersaturation $c_\infty(t)/c_0$ in the bulk is respectively shown
in Figure \ref{r}, \ref{fig5}, and \ref{f}.

Kirchner \cite{kirchner1973} and Burton and Speight \cite{burton1986}
have modeled the coarsening kinetics for prismatic dislocation loops
(KBS model).
Their model is based on the line tension approximation for the loops
and uses the same assumptions as in our DD simulations: loops cannot
glide and only vacancy bulk diffusion makes the loops grow or shrink.
A good agreement has been observed with experimental data 
\cite{powell1975,liu1995,bonafos1998} when these assumptions are valid. 
We check, in the present work, if such an agreement can also be obtained
with our DD simulations.

According to this coarsening model, the average radius of the loops $R_{\rm av}(t)$
should increase like $t^{1/2}$, following the law
\begin{eqnarray}
R_{\rm av}(t) = R_{\rm av}(0) [1 + \alpha t]^{1/2},
\label{ravt}
\end{eqnarray} 
where the temperature dependent growth rate is given by 
\begin{eqnarray}
\alpha = \frac{\eta}{2R^2_{\rm av}(0)}\frac{c_0 D_{\rm v}\mu\Omega}{kT}.
\label{eq:alphaBS}
\end{eqnarray}
Here $\eta$ is a geometric factor depending 
on the approximation used and on the boundary conditions 
when solving the vacancy diffusion equation for an isolated 
loop. 
Burton and Speight \cite{burton1986} 
assumed that the loops are acting as a point source / sink 
for vacancies and obtained $\eta=2$. 
This value has been found in good agreement 
with the one derived from our DD simulations of the shrinkage
of an isolated loop (\cf Appendix \ref{sec:isolated_loop}). 
We therefore use this value.
Using in Eq. \ref{eq:alphaBS} the input parameters of our DD simulations, 
we compare the time evolution of the loop average radius observed 
in our DD simulations with the one predicted by Eq. (\ref{ravt}).
Figure \ref{r} shows that a quantitative agreement is obtained
for all the initial loop areas studied.
As predicted by the KBS model, the average size of the loops only depends 
on its initial value and not on the loop density.

The KBS model predicts that the number of loops $N_{\rm loops}(t)$ 
decreases like the inverse of the time:
\begin{eqnarray}
	N_{\rm loops}(t) &=& \frac{N_{\rm loops}(0)}{1 + \alpha t}.
	\label{nt}
\end{eqnarray} 
This is in perfect agreement with the results obtained in our DD 
simulations (Figure \ref{fig5}).

The vacancy supersaturation is directly linked to the average
size of the loops according to the KBS model:
\begin{equation}
\frac{c_\infty(t)}{ c_0} = 1 + \frac{\mu b \Omega}{kTR_{\rm av}(t)}.
\label{cc0}
\end{equation} 
It should therefore decay asymptotically like $t^{-1/2}$.
The agreement with our DD simulations is not as perfect as for $R_{\rm av}(t)$
and $N_{\rm loops}(t)$, but the discrepancy is nevertheless small (Figure \ref{f}).
This difference may arise from the line tension approximation
used in the KBS coarsening theory
where it is assumed that a loop of radius $R$ is subject to the stress
$\tau=\mu b / R$. 
A more precise expression of the line tension exist 
for a circular prismatic loop \cite{lothe}. 
Such a loop of radius $R$ is indeed subject to the self stress
\begin{equation}
  \tau(R) = \frac{\mu b}{4\pi(1-\nu)R} \log{\left( \frac{4 R}{r_{\rm c}} \right)}.
  \label{eq:line_tension}
\end{equation}
The core radius $r_{\rm c}$ appearing in this expression is directly linked 
to the parameter $a$ used to spread dislocation core in the non singular
expressions of the stress field \cite{cai2006}.
For a prismatic loop, one should take $r_{\rm c}=a/2$, \ie 1.5\,{\AA} in our case.
With this value of the core radius, 
$\tau(R)R/\mu b$ varies between 0.96 and 1.08
for loop radii between 100 and 250\,nm. 
This shows the correctness of the approximation used by Kirchner \cite{kirchner1973}
and Burton and Speight \cite{burton1986} for the line tension. 
This small difference on the value of the line tension only slightly impact
the vacancy supersaturation. 
Using Eq. (\ref{eq:line_tension}) for the line tension instead of $\mu b / R$, 
the time evolution of the vacancy supersaturation is given by
\begin{equation}
  \frac{c_\infty(t)}{ c_0} = 1 + \frac{ \mu b \Omega}{4 \pi (1-\nu) kT R_{\rm av}(t)} 
  	\log{\left( \frac{4 R_{\rm av}(t)}{r_{\rm c}} \right)}.
  \label{eq:cc0_KBS_improved}
\end{equation}
Figure \ref{f} shows that the change on the vacancy supersaturations is small
in the considered size range.
This does not really improve the agreement with our DD simulations.
Probably, even Eq. (\ref{eq:line_tension}) based on an improved line tension approximation
only roughly estimates the stress existing on the dislocation segments in the DD simulations.
Eq. (\ref{eq:line_tension}) assumes that the loops are circular and neglect 
the interaction between different loops: DD simulations do not make these approximations.
On the other hand, one cannot exclude that the discretization of the loops in our DD simulations
induces a noise on the line tension which may impact the vacancy supersaturation. 
For instance, 5 Gauss points on each segment were used to integrate forces on each segment
so as to obtain nodal forces. A better estimation of the line tension may require more Gauss
points or an analytical expression of the nodal forces \cite{arsenlis2007}.
Despite all these limitations on the comparison between the results of our DD simulations
and the predictions of the KBS model, the agreement on the vacancy supersaturation
is reasonable. 

Finally, the KBS theory predicts that the size distribution of the loops,
once normalized, is stationary and is given by
\begin{eqnarray}
g(z) = \frac{1}{8{\rm e}^2}\frac{z}{(z-2)^4}
\displaystyle\exp\left( \frac{4}{z-2} \right )
\label{gz}
\end{eqnarray} 
if $0 < z < 2$, and 0 otherwise. 
$z=R/R_{\rm av}(t)$ is a normalized radius and $g(z)\md{z}$
is the probability to find a loop with a normalized size between $z$ 
and $z+\md{z}$.
The normalized size distribution in our DD simulations is stationary
and perfectly obeys Eq. (\ref{gz}), as can be seen in Figure \ref{figdistrib}.

A perfect agreement is therefore obtained between the simulations 
and the coarsening model for prismatic loops of Kirchner \cite{kirchner1973}
and Burton and Speight \cite{burton1986}. The agreement shows that this model 
is well suited when studying the coarsening of loops by vacancy bulk diffusion.
%This also allows us to conclude that the discretization of the dislocation lines
%in our simulations does not introduce any artifact. 

\section{Contribution of glide to the loop coarsening}

The advantage of DD simulations is that they are not restricted
to the study of climb associated with bulk diffusion.
We can superpose dislocation glide to see how it affects loop coarsening.
Loops are able to glide if they are unfaulted. 
One observes then experimentally \cite{eyre1971} 
a faster coarsening kinetics than with only bulk diffusion.

\subsection{Combining climb and glide motion}

The drag coefficient, the inverse of the mobility, for dislocation glide
is mainly controlled by phonon drag in pure fcc metals and varies linearly
with the temperature.
The temperature dependence in aluminum  was obtained by 
molecular dynamics calculations by Kuksin and coworkers \cite{kuksin} 
who found 
\begin{eqnarray}
  B_{\rm gl}(T) = B_{\rm gl}(\mathrm{300\,K}) \frac{T}{300},
	\label{eq:drag_glide}
\end{eqnarray} 
where $B_{\rm gl}(\mathrm{300\,K}) = 1.4\times 10^{-5}$\,Pa$\cdot$s 
is the value of the drag coefficient at temperature 300\,K. 
This value lies in between the two available experimental data 
$B_{\rm gl}(\mathrm{300\,K}) \approx 0.6\times10^{-5}$\,Pa$\cdot$s 
(Ref. \cite{hikata}) 
and $B_{\rm gl}(\mathrm{300\,K}) \approx 2.6\times 10^{-5}$\,Pa$\cdot$s 
(Ref. \cite{gorman}), and is also close 
to other molecular dynamics simulation results by Olmsted and coworkers \cite{olmsted},
$B^{\rm edge}_{\rm gl} = 1.2\times10^{-5}$\,Pa$\cdot$s 
and $B^{\rm screw}_{\rm gl} = 2.2\times10^{-5}$\,Pa$\cdot$s at 300\,K, 
and Groh and coworkers \cite{groh}, 
$B_{\rm gl}(\mathrm{300\,K}) = 4.5\times10^{-5}$\,Pa$\cdot$s. 
For simplicity, we neglect in our simulations 
that screw segments glide slower than edge 
segments and we consider a drag coefficient for glide which does not depend on the 
dislocation character, as given by Eq. (\ref{eq:drag_glide}).

The important point is that the glide mobility is much higher than the climb
mobility at the considered temperatures.
At 600\,K, the glide mobility is $M_{\rm gl}=3.3\times10^5$\,Pa$^{-1}\cdot$s$^{-1}$,
whereas the climb mobility for edge dislocations is only 
$M_{\rm cl}=1.75\times10^{-5}$\,Pa$^{-1}\cdot$s$^{-1}$.
Due to this difference of 10 orders of magnitudes between both mobilities,
it is not possible to handle both dislocation motions in the same step in 
our DD simulations.
The time interval compatible with the glide mobility 
would be so small that no climb could be observed during this period.

We therefore perform the glide and climb motions separately with two different
time steps, using an adiabatic approximation which assumes that the
degrees of freedom corresponding to dislocation glide
reach an equilibrium between two successive climb events.
The dislocation microstructure is equilibrated first
with respect to the glide motion.
Once glide is not producing any plastic strain, one climb step is 
performed: the glide mobility is put to zero and the time step 
is set to a value compatible with the climb mobility.
After this climb event, we equilibrate again the dislocation microstructure
with respect to glide and then go back to climb, thus cycling between climb events
and glide equilibration.
The dislocations are  considered to be in equilibrium with 
the glide motion, when the relative difference in the change of the 
Frobenius norm of the plastic strain tensor $\varepsilon$ between two 
consecutive time steps is less, than $10^{-3}$. Assuming index notation 
this can be expressed as
\begin{eqnarray}
\left | 1 - \frac{\sqrt{\varepsilon_{ij}(t+{\rm d}t) \varepsilon_{ij}(t+{\rm d}t)}}{\sqrt{\varepsilon_{ij}(t)\varepsilon_{ij}(t)}} \right | < 10^{-3}.
\end{eqnarray} 
We check that the result of our simulations does not depend
on the value of this threshold by performing some simulations
also with a $10^{-4}$ threshold.

Finally, one should stress that the 
definite $\{111\}$ initial glide planes are lost in our DD simulations.
A climbing dislocation is jogged. It therefore, 
does not lie, on average, in a definite $\{111\}$ plane.
The nodal representation of the dislocation lines does not describes
all jogs existing on the dislocation, but ``coarse-grains'' the line
to use less nodes per dislocation length.
As a consequence, the dislocation line vector $\mathbf{\zeta}$
may lie in a plane different from a $\{111\}$ plane. 
The dislocation glide plane is then fixed by this line vector $\mathbf{\zeta}$
and the Burgers vector $\mathbf{b}$.
The mobility of the jogged dislocation should be smaller than the one 
of the perfect dislocation as revealed by atomic simulations \cite{Rodney2000}.
Nevertheless, this is negligible compared to the mobility difference between
glide and climb. We therefore do not consider the effect of jogs on the glide mobility.
This is different from the picture of a jogged dislocation
in DD simulations where dislocations are discretized in edge and 
screw segments \cite{mordehai2008,mordehai2009}: 
in such simulations, all segments belong to a definite 
$\{111\}\langle110\rangle$ glide system, but the jogs 
have to be concentrated in a single point instead of being
spread all along the dislocation line, thus creating 
a superjog in the simulation.

\subsection{Glide and loop coarsening}

\begin{figure}[!bthp]
\begin{center}
  \frame{\includegraphics[angle=0,width=0.25\textwidth]{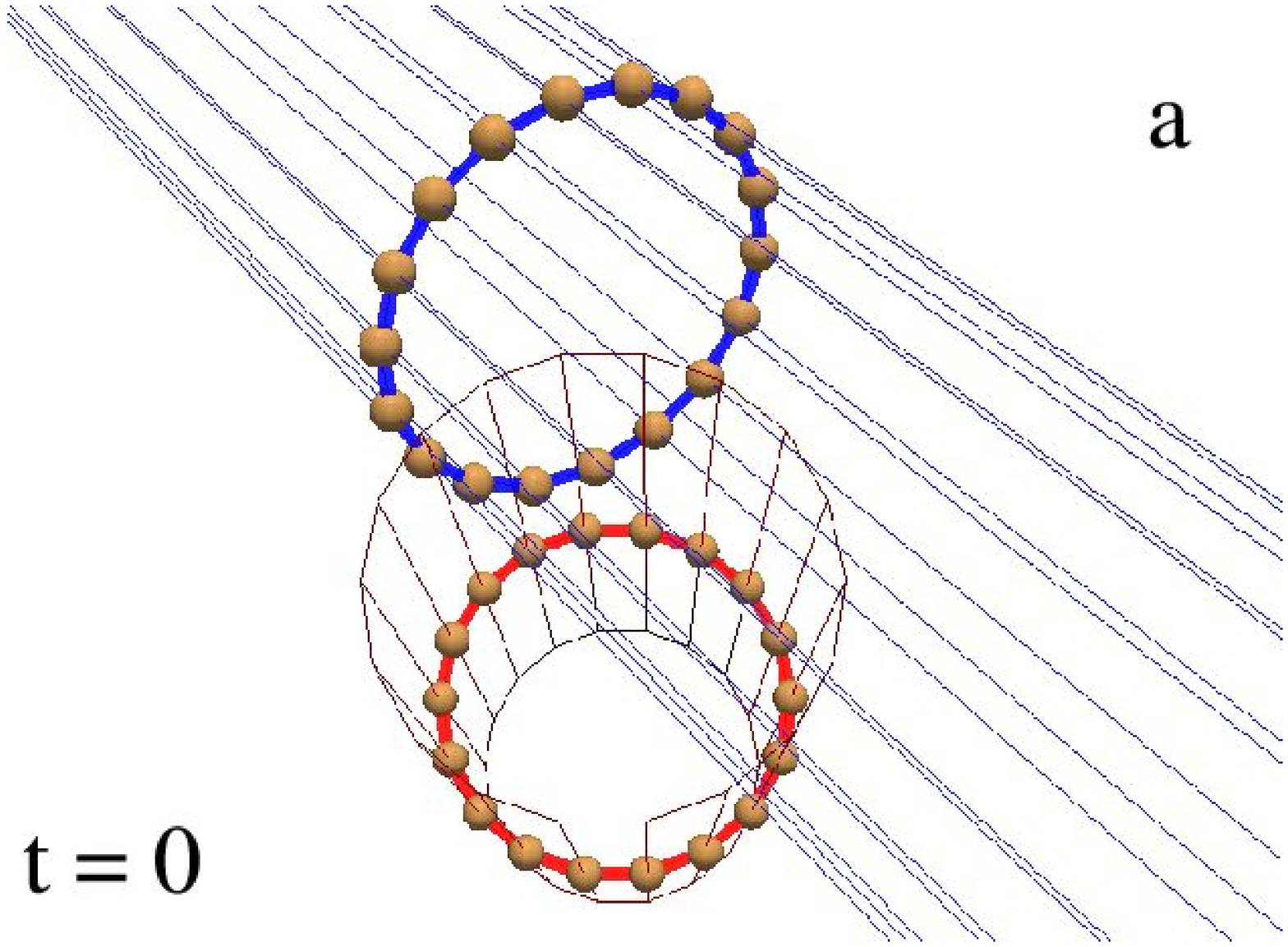}}
  \frame{\includegraphics[angle=0,width=0.25\textwidth]{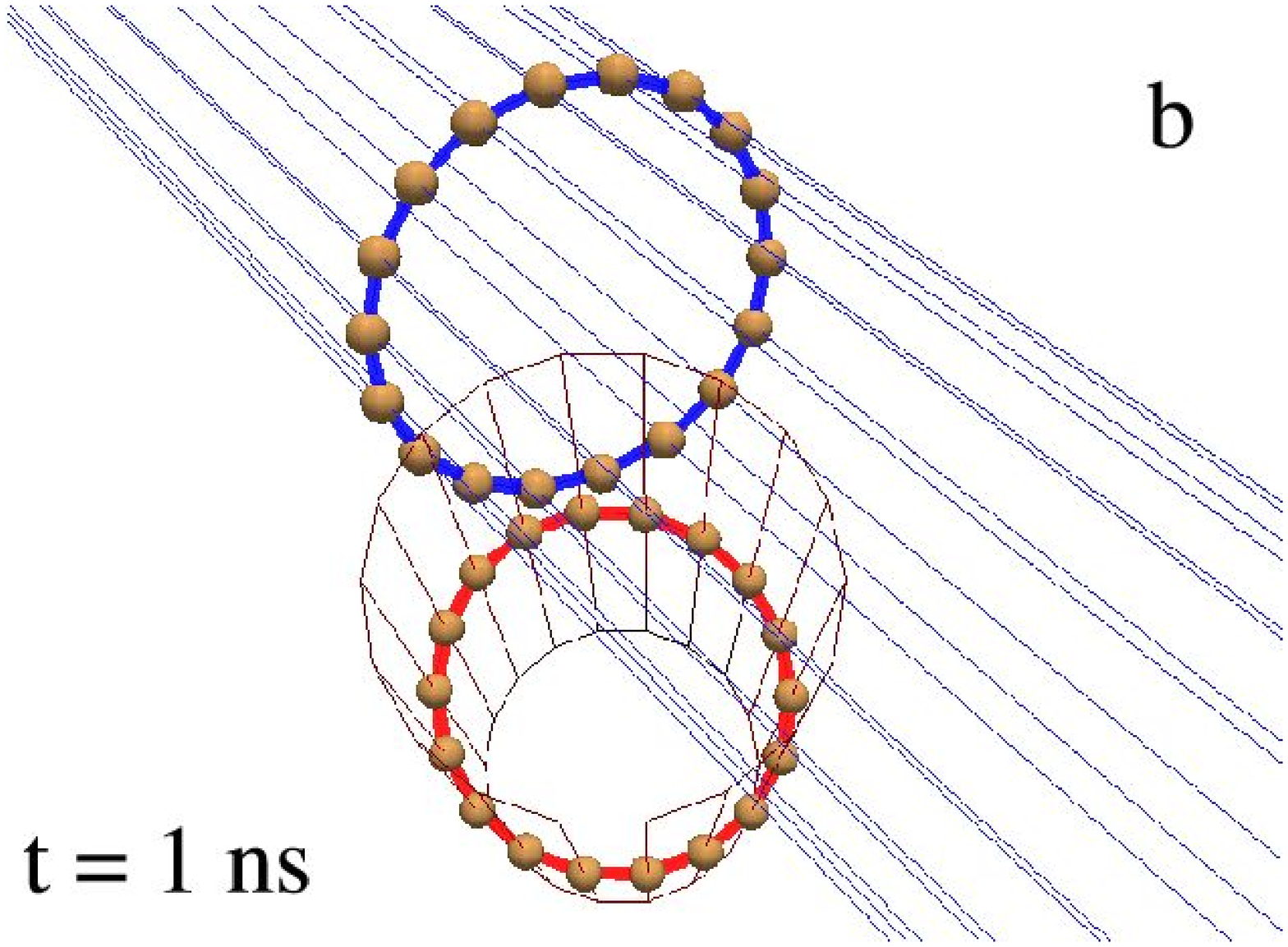}}
  \frame{\includegraphics[angle=0,width=0.25\textwidth]{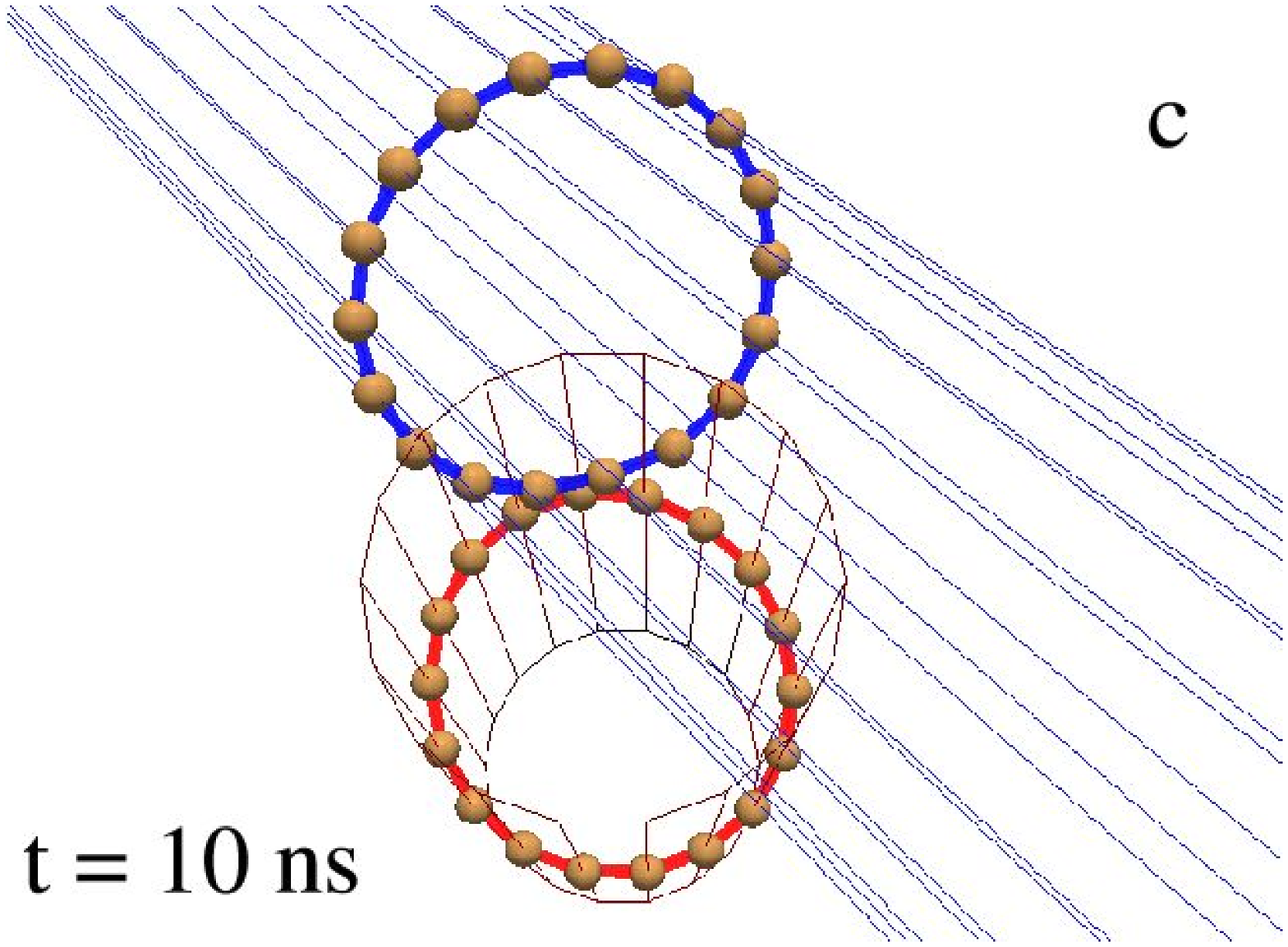}}\\
  \frame{\includegraphics[angle=0,width=0.25\textwidth]{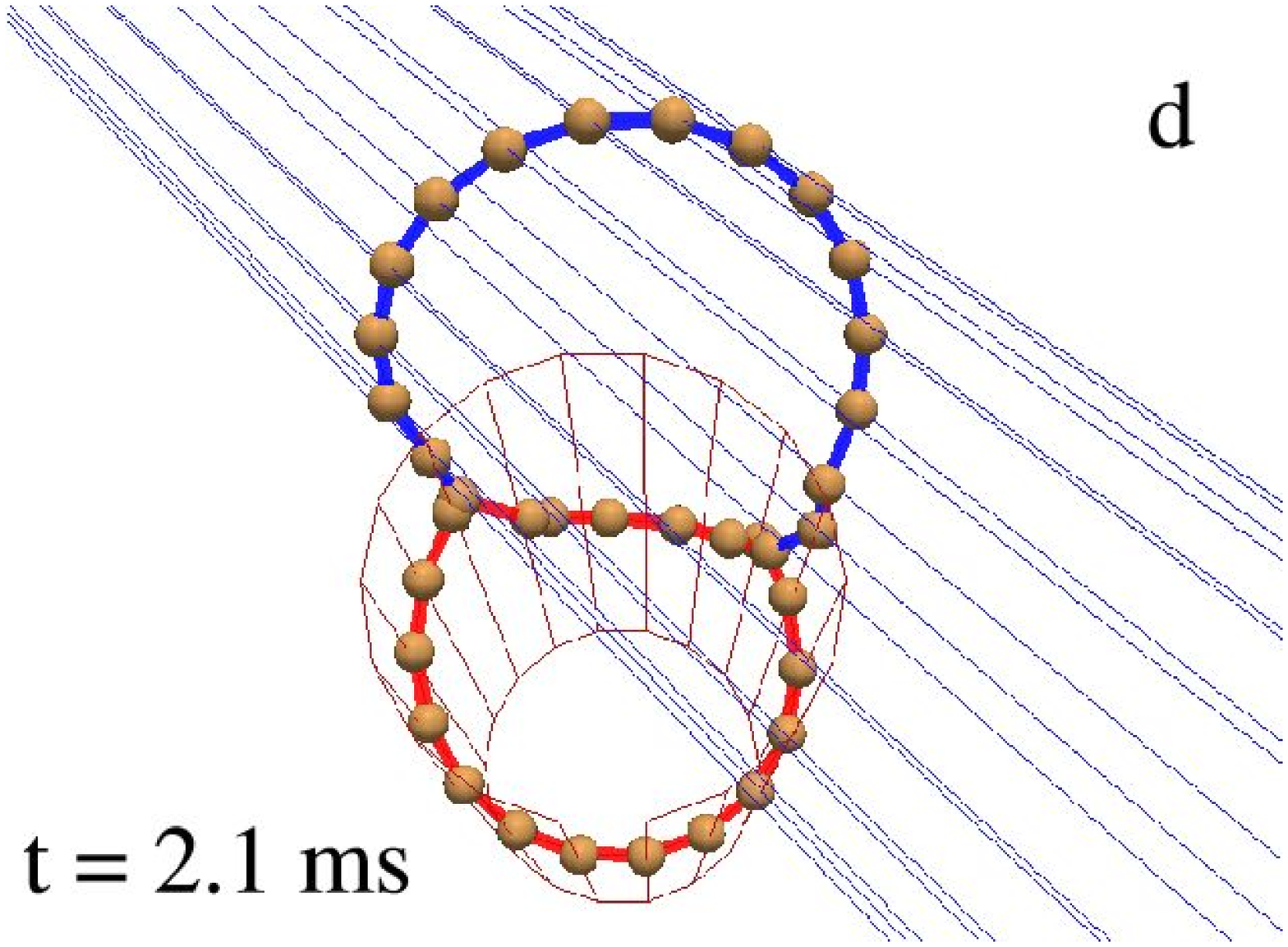}}
  \frame{\includegraphics[angle=0,width=0.25\textwidth]{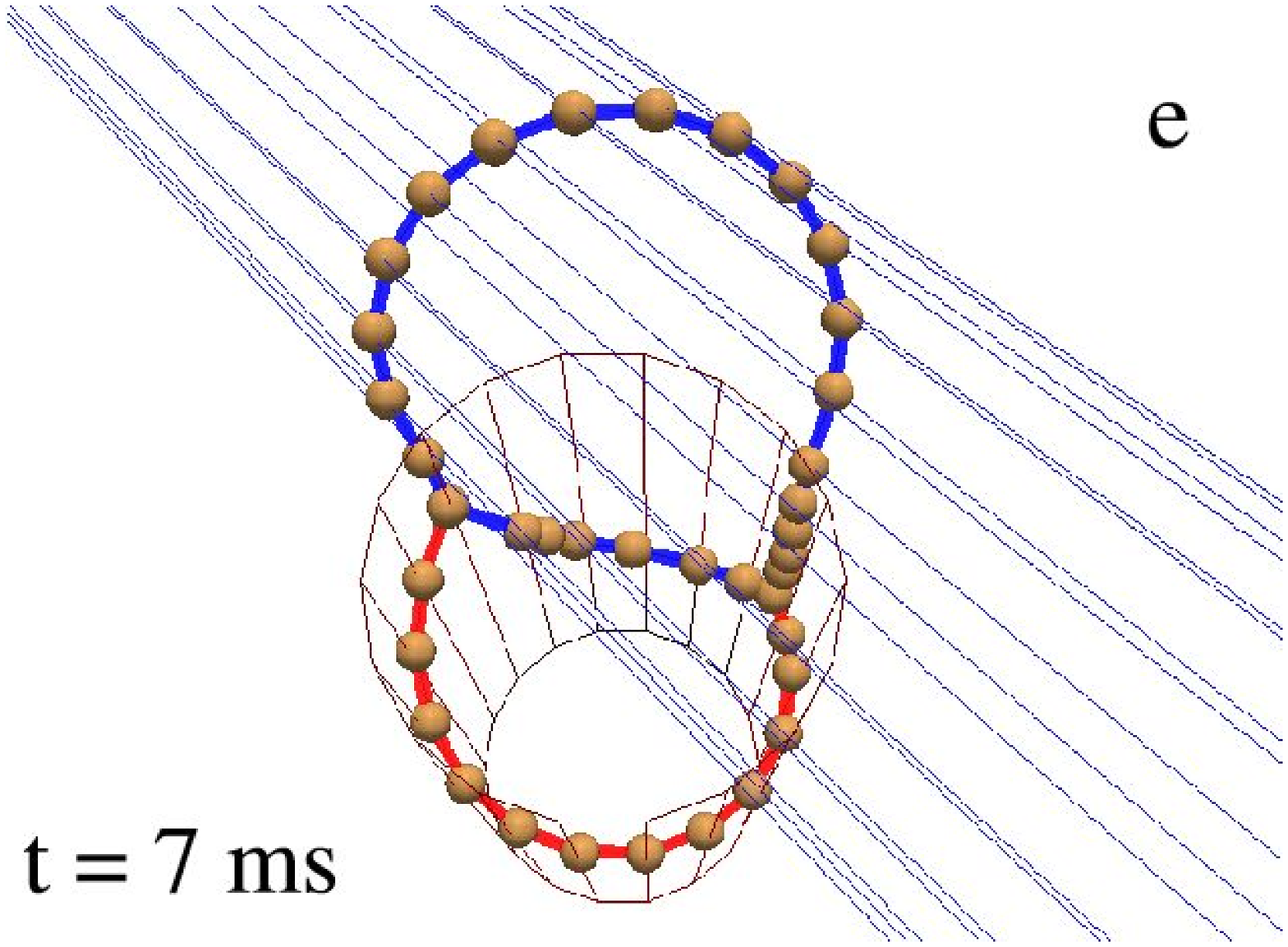}}
  \frame{\includegraphics[angle=0,width=0.25\textwidth]{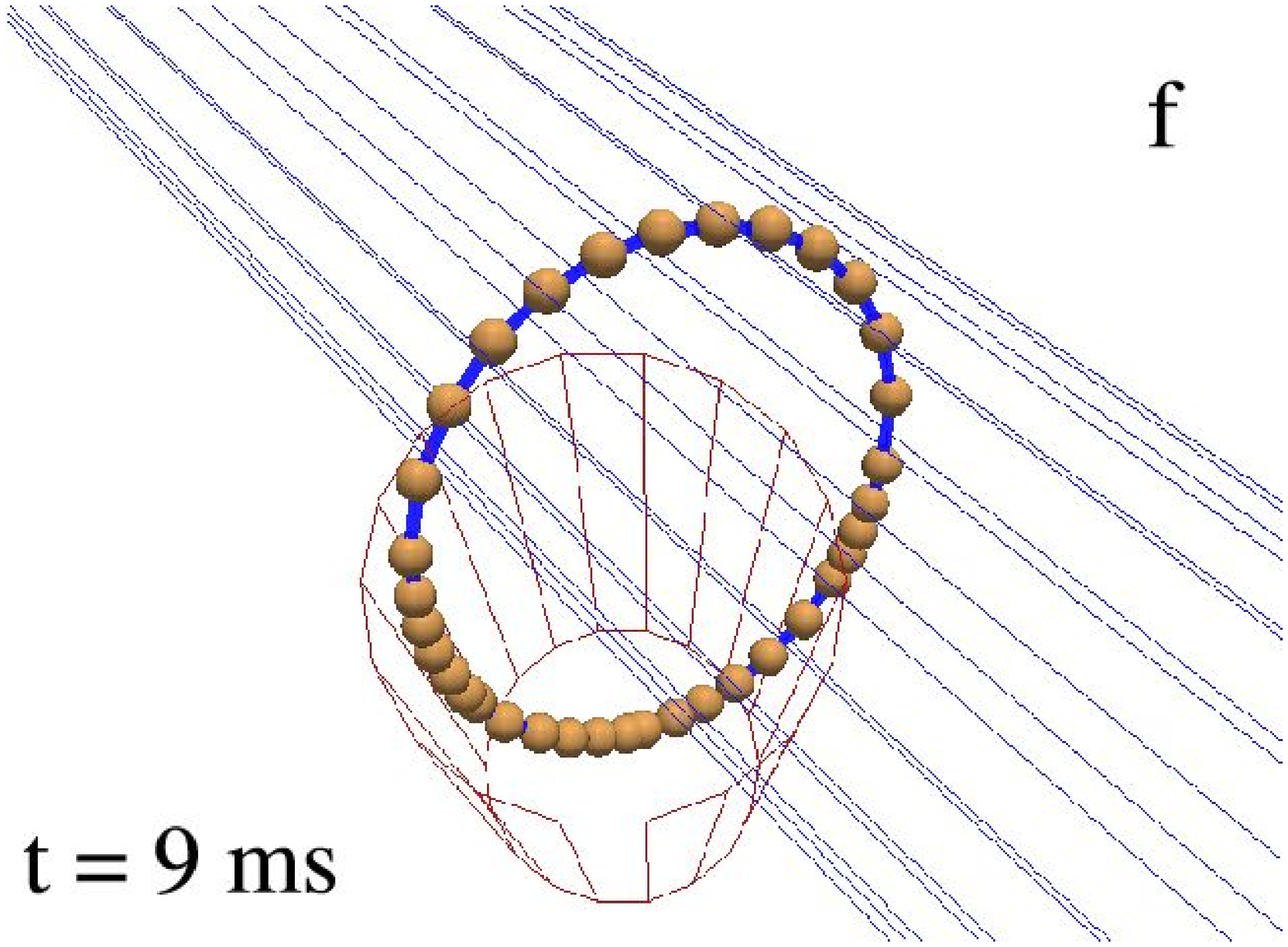}}
\end{center}
\caption{[Color online] Coarsening of a system formed by two attracting 
loops gliding on prismatic cylinders which intersect each other.
The initial glide cylinders of both loops are sketched with thin lines.
The axis of these cylinders correspond to the Burgers vector of the loops
which are $a/2 [101]$ and $a/2[\bar{1}10]$ respectively for the red 
and the blue loops.
The Burgers vector of the junction that can be seen in the insets d and e 
is $a/2[011]$
The initial radius of the red loop is $r=210$\,nm 
and of the blue loop $r=230$\,nm.}
\label{gc1}
\end{figure}

When the glide mobility is non-zero, prismatic loops can glide on the 
surface of a cylinder, whose axis is parallel to the Burgers vector of 
the loops. 
Due to the elastic interactions between them, 
loops move on this prismatic cylinders. 
The first consequence is that they can deviate now from 
their pure edge orientation.
Most importantly, if the cylinders of two different loops intersect,
the loops can come into contact with each other.
Loops can thus merge by gliding on their prismatic cylinder, 
a process much faster than coarsening by bulk diffusion. 
This is illustrated in Figure \ref{gc1} which shows the time
evolution of a system formed by two attracting prismatic loops.
The two circular loops at $t=0$ (Figure \ref{gc1}a) approach each other in a 
short time by gliding on their prismatic cylinders (Figure \ref{gc1}b)
until they come in contact and reach an equilibrium configuration
(Figure \ref{gc1}c). 
They then climb, the largest loop absorbing the vacancies emitted by
the smallest one (Figures \ref{gc1}d-\ref{gc1}f).
During all this coalescence stage, glide allows the loops to keep
their equilibrium shape.
At the end (Figure \ref{gc1}f), only the largest loop survives.
The coarsening thus results from climb assisted by glide.
One should also note that, in this case, the loops are so close 
that the mechanical climb force does not only arises from the loop
line tension but also from the stress exerted from one loop to the other one.
This is another factor, associated with loop glide, which leads
to a speed-up of the coarsening.

\begin{figure}[!tbhp]
\begin{center}
\includegraphics[angle=270,width=0.75\textwidth]{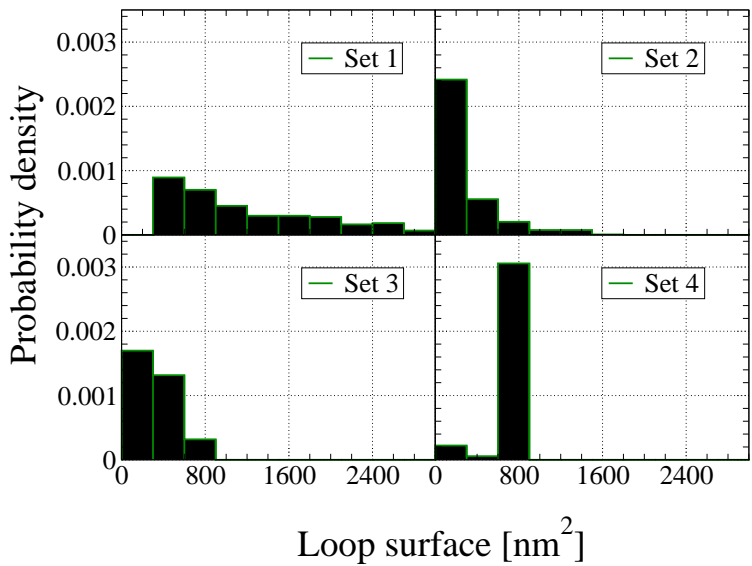}
\end{center}
\caption{Initial probability density distribution of loop surfaces for the 
simulations with glide and climb.}
\label{figloopsinitdistrib}
\end{figure}

\begin{figure}
\begin{center}
\frame{\includegraphics[angle=0,width=0.40\textwidth]{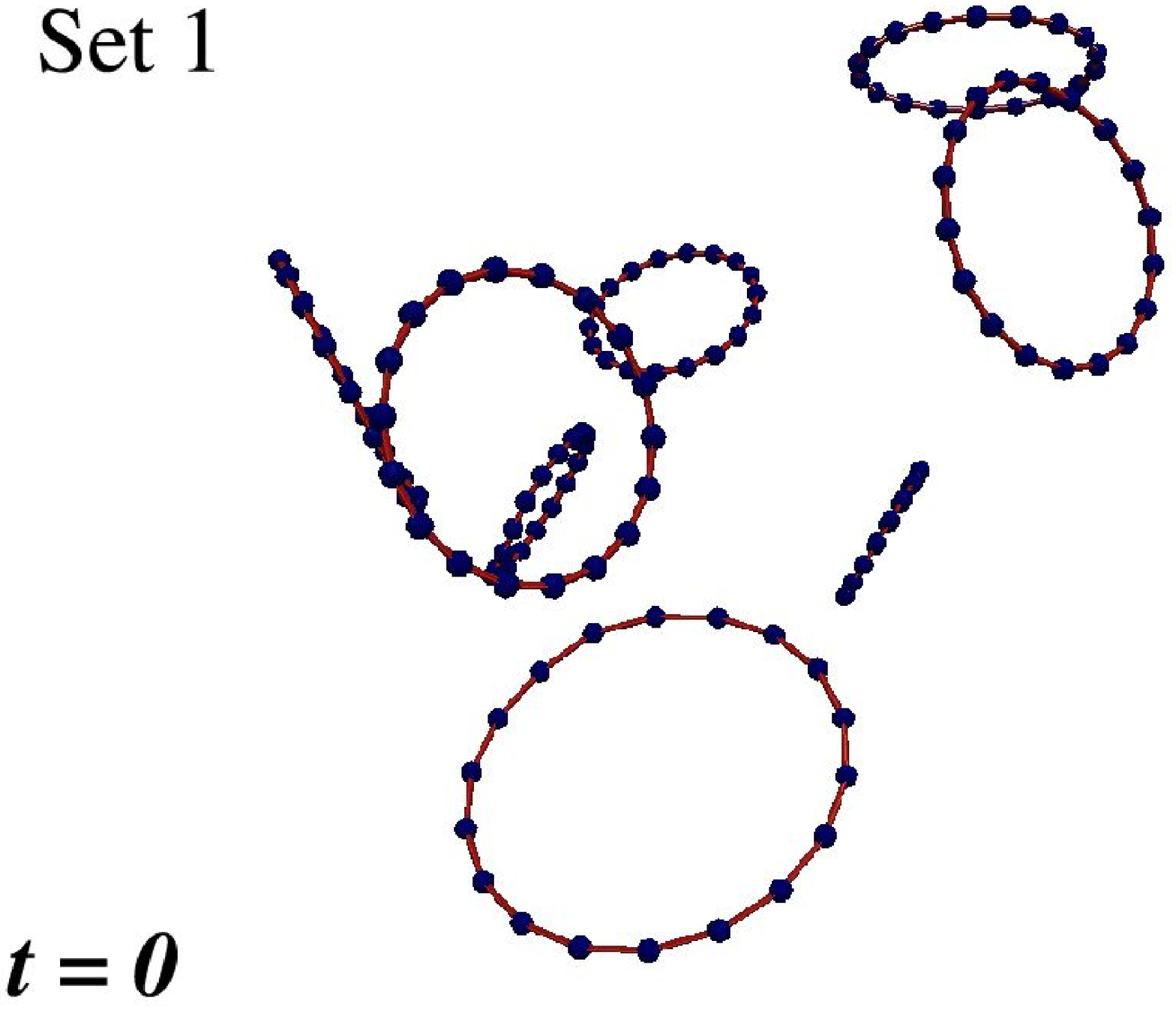}}
\frame{\includegraphics[angle=0,width=0.40\textwidth]{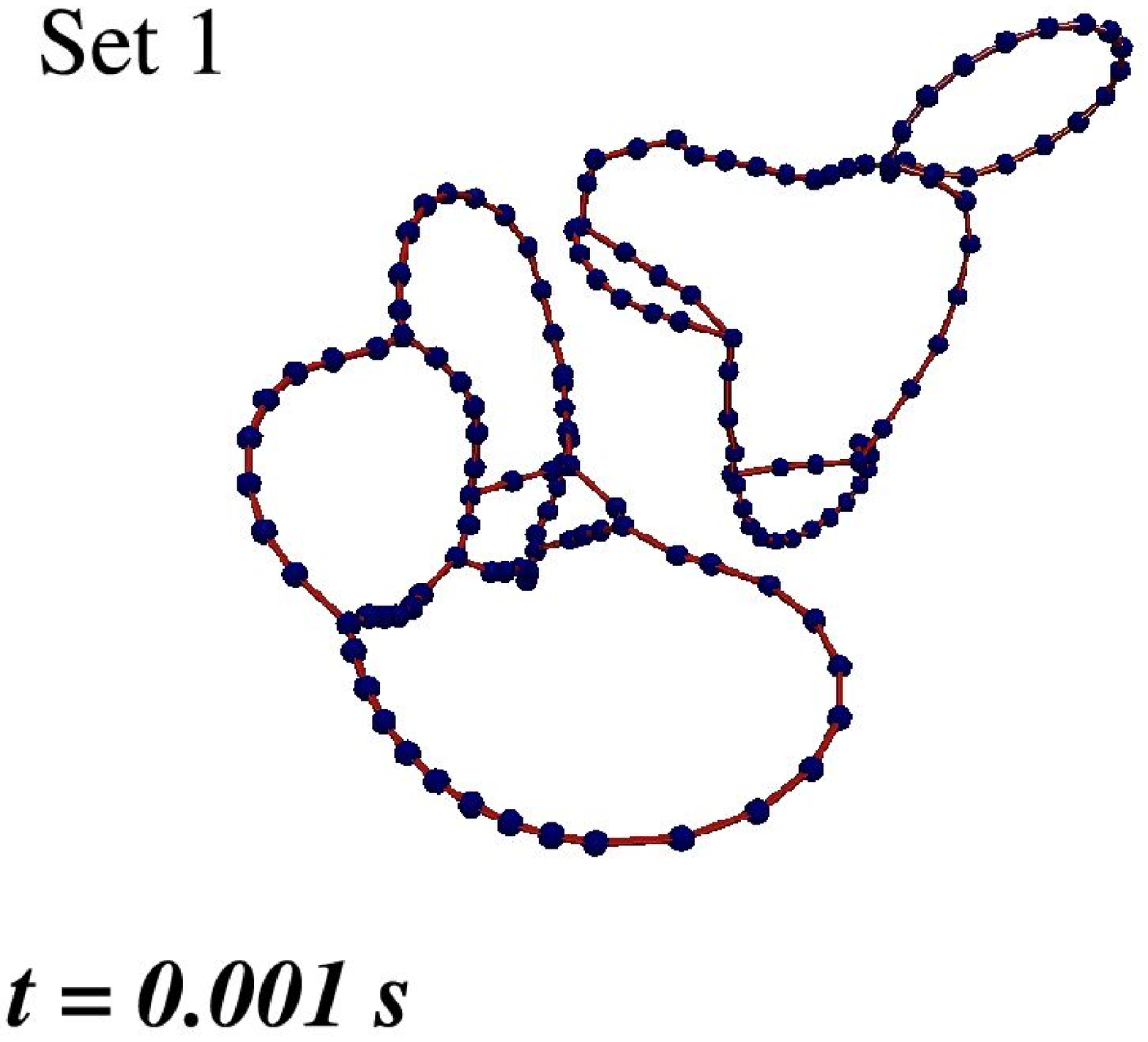}}\\
\frame{\includegraphics[angle=0,width=0.40\textwidth]{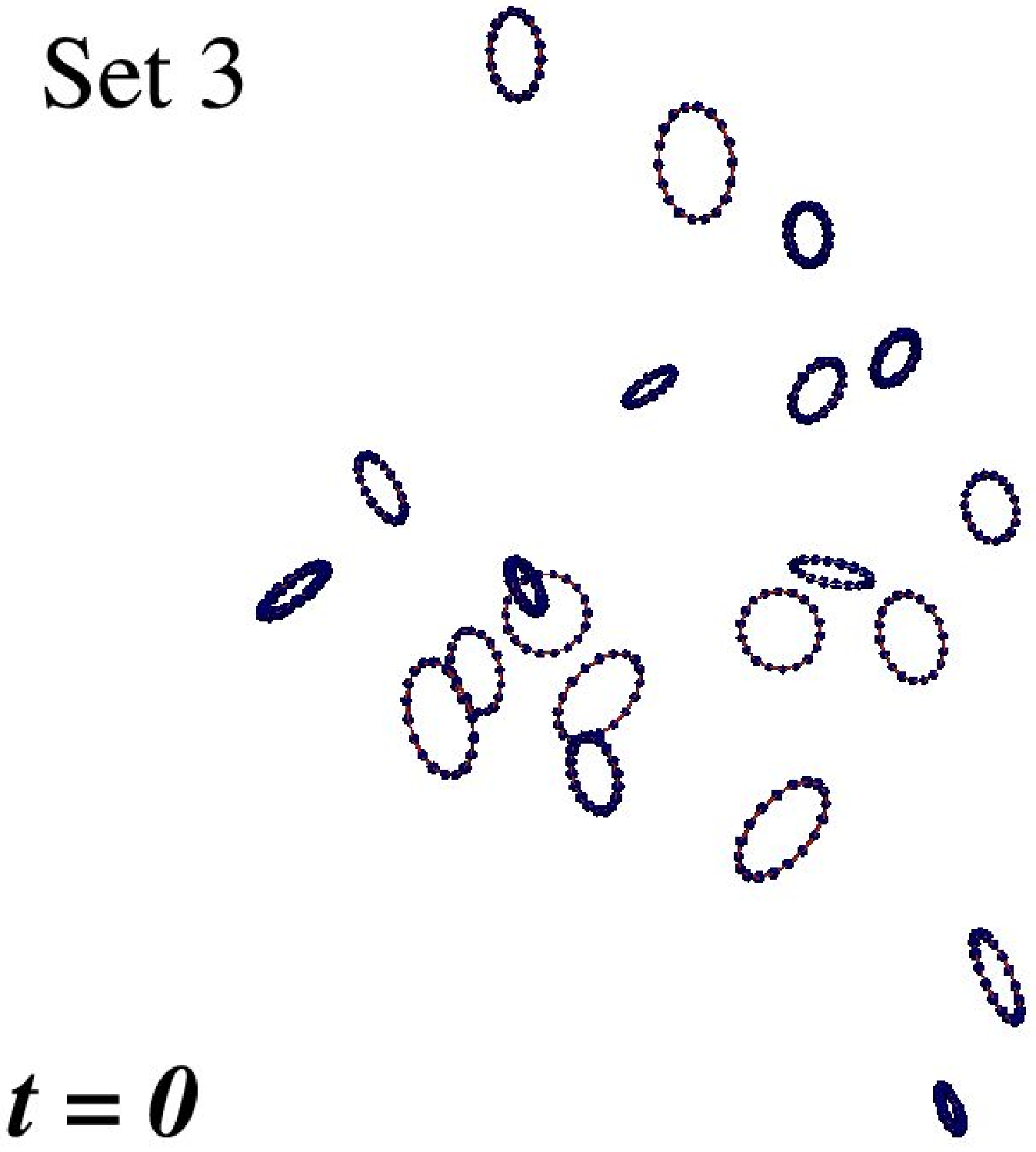}}
\frame{\includegraphics[angle=0,width=0.40\textwidth]{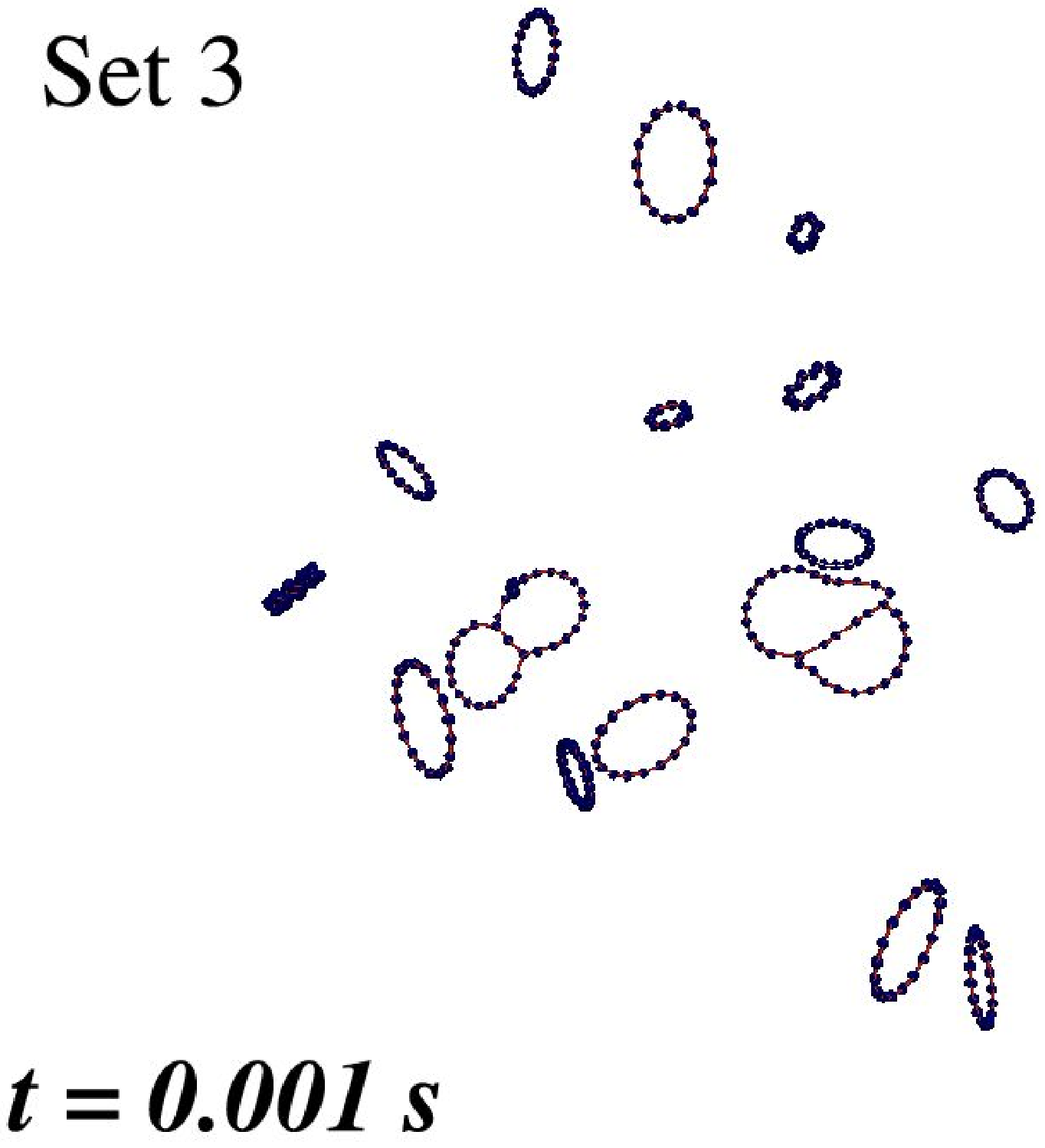}}
\end{center}
\caption{[Color online] Coarsening dynamics of loops in a typical 
simulation for Set 1 (top) and Set 3 (bottom).
(See video as supplementary online material.)}
\label{coalescence}
\end{figure}

\begin{figure}
\begin{center}
\includegraphics[angle=270,width=0.75\textwidth]{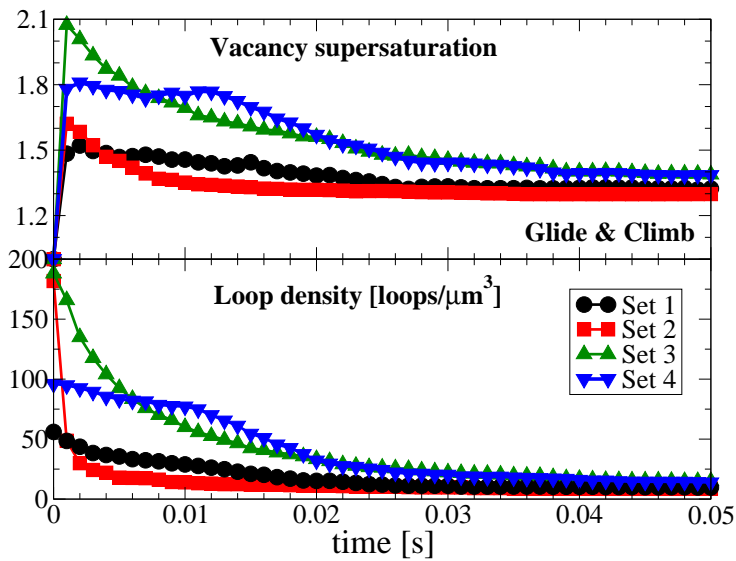}
\end{center}
\caption{[Color online] Time evolution of the vacancy supersaturation
$c_\infty(t)/c_0$ 
and of the loop density in the simulations with glide and climb.}
\label{fig_q_of_int_gc}
\end{figure}

The influence of the glide on the time evolution of the quantities of 
interest (vacancy supersaturation, loop density, mean projected loop 
area) was studied on four different sets of loops. 
The initial size distributions corresponding to the four different 
sets of simulations are presented in Figure \ref{figloopsinitdistrib}.
All sets contain the same number of vacancies condensed in the loops,
but some sets correspond to a high density of small loops 
(sets 2 and 3), whereas some other sets to a broader distribution
with larger loops and a smaller density (set 1).
The side of the simulation box was set to $L = 500$\,nm
and the statistics was obtained by averaging over 50 statistically
equivalent simulations.
Initially no vacancy supersaturation exists in the simulation box
($c_{\infty}(t=0)=c_0$).
The area of the loops was calculated by projecting the loops on planes 
perpendicular to the Burgers vector of the loops. It gives thus a measure
of the number of vacancies condensed in the loops.

Glide makes the loop population rapidly evolve at the beginning
of the simulation. 
In set 1, which contains large loops close to each other,
loops come in contact by glide and form a complicated network 
(Figure \ref{coalescence}) which then evolves by climb assisted by glide.
In contrast, in set 3, which is a collection of small loops separated
by larger distances compared to their radius, few loops coalesce by glide
at the beginning. The coarsening mainly proceed by climb, with glide
leading to some isolated coalescence events and thus enhancing the kinetics.

The time evolution of the vacancy supersaturation and of the loop 
densities is presented in Figure \ref{fig_q_of_int_gc}. 
The vacancy supersaturation initially increases for short time 
so as to reach equilibrium with the given loop population. 
This is similar to what was observed in the previous section without glide.
Then the supersaturation decays as the loops absorb the excess of vacancies 
during the coarsening, whereas the loop density decreases 
and their size increases.

\begin{figure}
\begin{center}
\includegraphics[angle=0,width=0.75\textwidth]{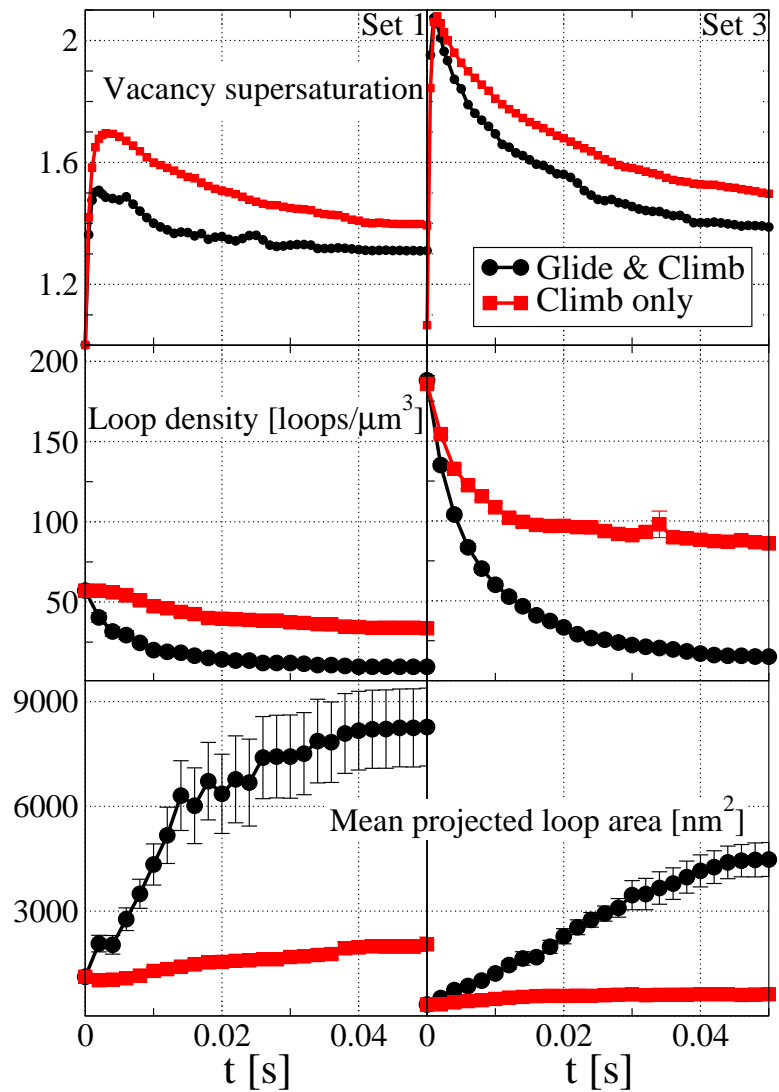}
\end{center}
\caption{[Color online] Time evolution of the vacancy supersaturation, 
loop density, and mean projected loop area in the simulations with 
finite glide drag coefficient, compared to the simulations with infinite 
glide drag coefficient, for the data set 1 (on the left) and data set 3 
(right column).}
\label{gc_and_c}
\end{figure}

To highlight the effect of glide on the coarsening of the 
prismatic loops, the simulations were repeated on the data 
sets 1 and 3, with a glide mobility set to zero.
As it can be seen from 
Figure \ref{gc_and_c}, the simulations with glide and climb result in 
faster coarsening than simulations with climb only.
Of course, when loops are allowed to glide, the coarsening model
of Kirchner \cite{kirchner1973} and Burton and Speight \cite{burton1986}
does not apply anymore. This is quite normal as this model assumes 
that loops can only climb thanks to vacancy bulk diffusion
and, as soon as loop glide is allowed, it dramatically 
changes the coarsening kinetics.

\section{Conclusion}

The introduction in DD simulations of a dislocation climb model
based on vacancy bulk diffusion allows us to study the coarsening
kinetics of prismatic loops. When loops cannot glide,
because they are faulted for instance, 
we obtain a perfect agreement between our simulations and 
the coarsening model of Kirchner \cite{kirchner1973}
and Burton and Speight \cite{burton1986}.
The average size of the loops increases with time $t$ like $t^{1/2}$, 
the loop density decreases like $1/t$, 
and the vacancy supersaturation decreases like $t^{-1/2}$.

When the loops can also glide on their prismatic cylinders, 
a much faster coarsening kinetics is obtained.
Prismatic glide leads to direct coalescence of the loops.
These coalescence events enhance the coarsening of the loop
distribution.
For high loop densities, where the distance between loops 
is small compared to their size, glide leads to a complex
dislocation microstructure and coarsening mainly proceeds 
by aggregation.

Only vacancy bulk diffusion has been taken into account in our simulations.
It has been proposed in the literature \cite{johnson1960}
that vacancy pipe-diffusion also leads to coalescence of the loops,
as a result of a motion of the loops in the plane perpendicular
to their prismatic cylinder.
The next step of this work will therefore be to include climb
associated with vacancy pipe diffusion so as to simulate
all coarsening regimes.
The effects of jogs on climb \cite{caillard} should also be considered so as to improve
the climb mobility law.

\section*{Acknowledgments} 
This work was supported by the European fusion
materials modeling program (EFDA MAT-REMEV).
The simulations were performed with the 3D DD code 
Numodis (CEA-CNRS, France). The authors are grateful 
to the other developers of the code, M. Fivel, E. Ferrie, 
and V. Quatella.

\appendix
\section{Isolated loop}
\label{sec:isolated_loop}

\begin{figure}[!thbp]
\begin{center}
\includegraphics[angle=270,width=0.75\textwidth]{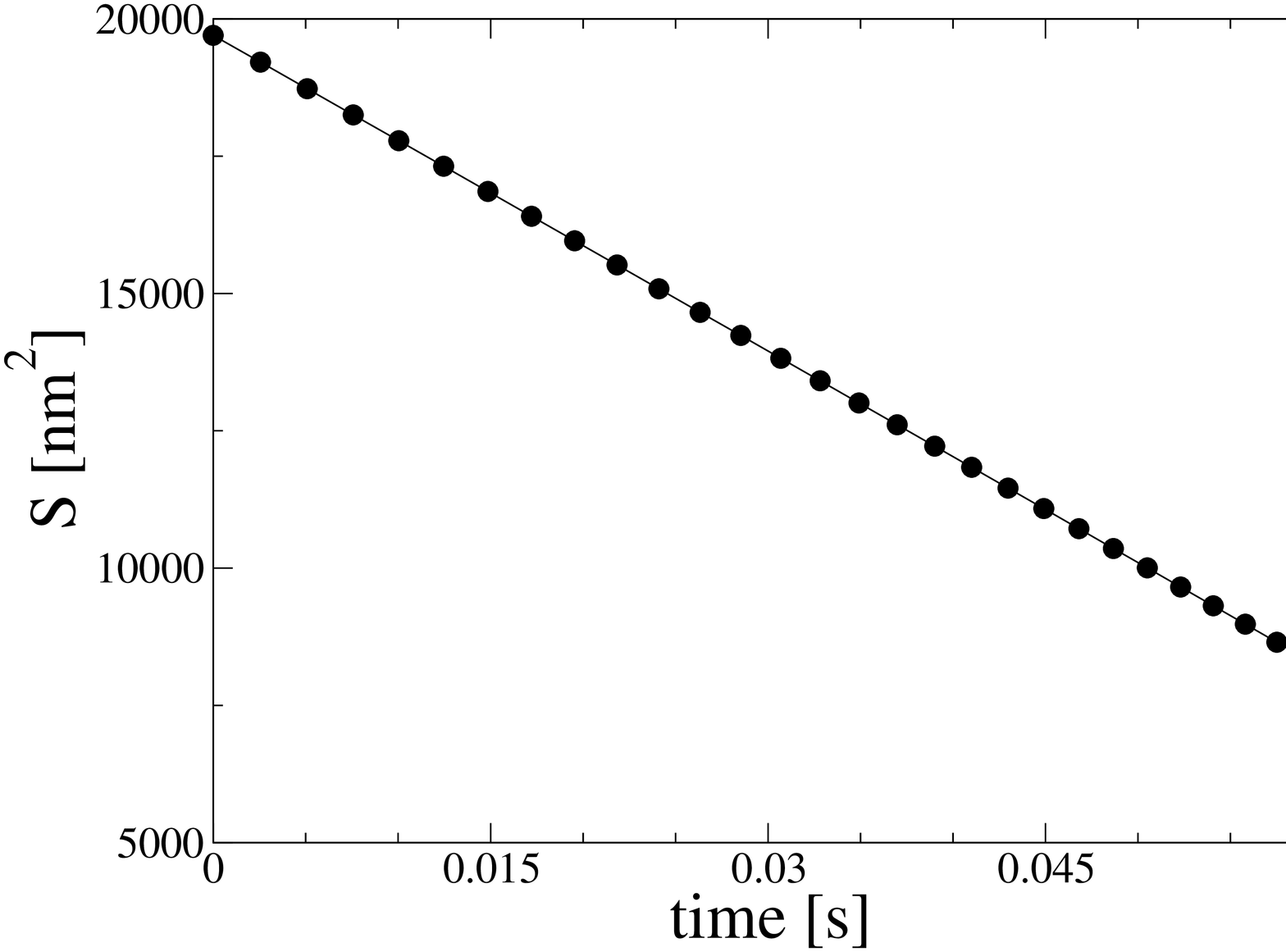}
\end{center}
\caption{Time evolution of the surface $S$ of an isolated loop 
with initial radius $R(0)=80$\,nm when $c_{\infty}(t)=c_0$.}
\label{ravts}
\end{figure}

The coarsening model of Kirchner \cite{kirchner1973}
and of Burton and Speight \cite{burton1986} makes use 
of the growing law of a prismatic loop. Such a law 
is based on the line tension approximation and 
on the solution of Fick's equation for vacancies 
diffusing from the bulk to the loop.
Several solutions can be found in the literature 
depending of the assumed geometry for the flux fields
and of the boundary conditions for the vacancy concentration.
These solutions differ only in the value of the geometrical 
factor $\eta$ which appears in Eq. (\ref{eq:alphaBS})
in the modeling of coarsening kinetics of prismatic loops.
Seidman and Balluffi \cite{balluffi} assuming toroidal boundary conditions 
for the flux field around the prismatic loop obtained 
$\eta(R) = \sqrt 6\pi / \ln{(8R/r_{\rm c})}$,
where $r_{\rm c}$ is the circular cross-section of the torus
and $R$ is its radius. 
If the loop is treated as a disc \cite{kirchner1973}, 
the value is $\eta = 4/\pi \sqrt{3/2} \approx 1.56$. 
Burton and Speight obtained $\eta = 2$ by 
assuming spherical symmetry for the vacancy flux field around a loop 
considered to be a punctual source/sink. 

Following the approach of Mordehai \etal \cite{mordehai2008},
we use our DD simulations in combination with a line tension 
model to evaluate the value of $\eta$.
We therefore simulate the loop shrinkage of an isolated loop
under equilibrium vacancy concentration $c_\infty = c_0$,
and we maintain fixed this vacancy concentration in the bulk.
One observes that a vacancy loop annihilates.
According to the line tension model 
\cite{kirchner1973,burton1986,balluffi,mordehai2008}
its surface $S$ is decreasing like 
\begin{equation}
	S(t) = S_0\left(1-\frac{t}{\tau_0}\right),
\end{equation}
where $S_0$ is the initial surface of the loop
and the annihilation time $\tau_0$ is given by
\begin{eqnarray}
	\tau_0 = \frac{k T S_0}{2 \pi \eta c_0 D_{\rm v} \mu  \Omega}.
	\label{eq:tau0}
\end{eqnarray}

The shrinkage of an isolated loop with initial radius $R(0)=80$\,nm is 
presented in Figure \ref{ravts}: the linear variation with 
time of the surface of the loop is clearly obtained.
Using Eq. (\ref{eq:tau0}), the value of the geometric factor
$\eta$ can be deduced.
For loops with radius from $R=80$\,nm up to $R=4$\,$\mu$m,
we get values $\eta \approx 1.92-1.98$, close to the value 
$\eta = 2$ obtained by Burton and Speight \cite{burton1986}. 
For the sake of simplicity we then also assume that $\eta(R)$ 
is constant, and we use the value $\eta = 2$.

%\bibliographystyle{tPHM}
%\bibliography{bako2011_PhilMag}

\end{document}